\documentclass[11pt]{article}
\usepackage{amssymb,graphicx}
\usepackage{rotating}

\usepackage{amsmath}
\usepackage[ansinew]{inputenc}
\usepackage{xcolor}

\usepackage[english]{babel}
\usepackage{helvet}
\usepackage{rotating}

\usepackage[comma]{natbib}

\usepackage{algorithm}
\floatname{algorithm}{Algorithm}

\newtheorem{prop}{Proposition}[section]

\begin{document}

%% Here are the title, author names and addresses
\title{A note on the consistency of the Narain-Horvitz-Thompson estimator}

\author{Guillaume Chauvet \thanks{ENSAI (CREST), Campus de Ker Lann, Bruz - France, chauvet@ensai.fr}}

\maketitle

\begin{abstract}
For the Narain-Horvitz-Thompson estimator to have usual asymptotic properties such as consistency, some conditions on the sampling design and on the variable of interest are needed. \citet{car:cha:gog:lab:2010} give some sufficient conditions for the mean square consistency, but one of them is usually difficult to prove or does not hold for some unequal probability sampling designs. We propose alternative conditions for the mean square consistency of the Narain-Horvitz-Thompson estimator. A specific result is also proved in case when a martingale sampling algorithm is used, which implies consistency under a fast algorithm for the cube method.
\end{abstract}


\newpage

\section{Introduction} \label{sec:1}

\noindent When a random sample $S$ is selected inside a finite population $U$, the Narain (1951)-Horvitz-Thompson (1952) \nocite{nar:1951,hor:tho:1952} estimator $\hat{t}_{y\pi}$ if often used for the total $t_y=\sum_{k \in U} y_k$ of some variable of interest. For the Narain-Horvitz-Thompson estimator to have usual asymptotic properties, such as asymptotic normality or consistency, some conditions on the sampling design and on the variable of interest are needed. Following the approach in \citet{rob:sar:1983} and \citet{bre:ops:2000}, \citet{car:cha:gog:lab:2010} give sufficient conditions for the mean square consistency. However, one of these conditions is related to the second-order inclusion probabilities and is usually difficult to prove for unequal probability sampling designs. \\

\noindent In this note, we propose alternative conditions for the mean square consistency of the Narain-Horvitz-Thompson estimator, i.e. under which
    \begin{eqnarray} \label{mean:square:cons}
      E\left\{N^{-1}(\hat{t}_{y\pi}-t_y)\right\}^2 & = & O(n^{-1})
    \end{eqnarray}
with $N$ the population size. The proposed conditions are usually easier to prove, and are known to hold for several sampling designs with unequal probabilities. We also give conditions under which the Narain-Horvitz-Thompson is consistent in mean square under a martingale sampling algorithm, which implies consistency under a fast algorithm for the cube method \citep{dev:til:2004}. Our asymptotic framework is that described in \citet{isa:ful:1982}. We assume that the population $U$ belongs to a nested sequence $\{U_t\}$ of finite populations with increasing sizes $N_t$, and that the population vector of values $y_{Ut}=(y_{1t},\ldots,y_{Nt})^{\top}$ belongs to a sequence $\{y_{Ut}\}$ of $N_t$-vectors. For simplicity, the index $t$ will be suppressed in what follows and all limiting processes will be taken as $t \to \infty$.

\section{Finite population framework} \label{sec:2}

\noindent We note $\pi=(\pi_1,\ldots,\pi_N)^{\top}$ a $N$-vector of probabilities. Let $p(\cdot)$ denote a sampling design in $U$ with parameter $\pi$, that is, such that the expected number of draws for unit $k$ in the sample equals $\pi_k>0$. Let $n=\sum_{k \in U} \pi_k$ denote the integer average sample size. A random sample $S$, with or without repetitions, is selected in $U$ by means of the sampling design $p(\cdot)$. The total $t_y$ is unbiasedly estimated by
    \begin{eqnarray} \label{est:ht:hh}
      \hat{t}_{y} & = & \sum_{k \in U} \frac{y_k}{\pi_k}~I_k,
    \end{eqnarray}
with $I=(I_1,\ldots,I_N)^{\top}$ and $I_k$ the number of times that unit $k$ is selected in the sample. The variance of $\hat{t}_{y}$ is
    \begin{eqnarray} \label{var:ht:hh}
      V \left(\hat{t}_{y}\right) & = & \sum_{k,l \in U} \frac{y_k}{\pi_k} \frac{y_l}{\pi_l}~Cov(I_k,I_l).
    \end{eqnarray}

\noindent If $p(\cdot)$ is a without-replacement sampling design, a same unit $k$ may appear only once in the sample and $I_k$ is a sample membership indicator. Formula (\ref{est:ht:hh}) yields the Narain-Horvitz-Thompson estimator $\hat{t}_{y\pi}$ whose variance is
    \begin{eqnarray} \label{var:ht}
      V \left(\hat{t}_{y\pi}\right) & = & \sum_{k \in U} \left(\frac{y_k}{\pi_k}\right)^2 \pi_k(1-\pi_k) + \sum_{k \neq l \in U} \frac{y_k}{\pi_k} \frac{y_l}{\pi_l} (\pi_{kl}-\pi_k \pi_l),
    \end{eqnarray}
with $\pi_{kl}$ the probability that units $k$ and $l$ are selected jointly in $S$. Poisson sampling \citep{haj:1964} is a particular without-replacement sampling design, obtained when the vector $I$ of sample membership indicators is obtained from $N$ independent Bernoulli trials. In such case, the variance of the Narain-Horvitz-Thompson estimator is
    \begin{eqnarray} \label{var:ht:pois}
      V\left(\hat{t}_{y\pi}^{po}\right) & = & \sum_{k \in U} \left(\frac{y_k}{\pi_k}\right)^2 \pi_k(1-\pi_k),
    \end{eqnarray}
which is the first term of the variance in (\ref{var:ht}) for any without-replacement sampling design. \\

\noindent If $p(\cdot)$ is a with-replacement sampling design, a same unit $k$ may appear several times in the sample and formula (\ref{est:ht:hh}) yields the \citet{han:hur:1953} estimator $\hat{t}_{yHH}$. Multinomial sampling is a particular with-replacement sampling design, obtained when the sample $S$ is obtained from $n$ independent draws, some unit $k$ being selected with probability $n^{-1} \pi_k$ at each draw. In such case, the variance of the Hansen-Hurwitz estimator is
    \begin{eqnarray} \label{var:ht:mult}
      V\left(\hat{t}_{yHH}^{mult}\right) & = & \sum_{k \in U} \pi_k \left(\frac{y_k}{\pi_k}-\frac{t_y}{n} \right)^2.
    \end{eqnarray}
With-replacement sampling designs are less common in surveys. We therefore confine our attention to without-replacement sampling designs and to the Narain-Horvitz-Thompson estimator $\hat{t}_{y\pi}$. However, the variance obtained under multinomial sampling will be a useful benchmark to prove the mean square consistency.

\section{Sufficient conditions for mean-square consistency} \label{sec:3}

\noindent From (\ref{var:ht}), we obtain
    \begin{eqnarray} \label{ineg:mean:square:1}
      V\left(\hat{t}_{y\pi}\right) & \leq & N^2 \left(\frac{1}{N~\min_{l \in U} \pi_l}+\frac{\max_{k \neq l \in U} |\pi_{kl}-\pi_k \pi_l|}{(\min_{l \in U} \pi_l)^2} \right) \times \frac{1}{N} \sum_{k \in U} y_k^2.
    \end{eqnarray}
This directly leads to Proposition \ref{prop0} below.

\begin{prop} \label{prop0} \citep{car:cha:gog:lab:2010}. Assume that the following conditions hold:
\begin{itemize}
  \item[] H1. We assume that $\lim_{t \rightarrow \infty} \frac{n}{N} = f \in ]0,1[$.
  \item[] H2. We assume that $\min_{k \in U} \pi_k \geq \lambda_1>0$.
  \item[] H3. The variable $y$ has a bounded second moment, i.e. there exists some constant $C_1$ such that $N^{-1} \sum_{k \in U} y_k^2 \leq C_1$.
  \item[] H4: We have $\limsup_{t \rightarrow \infty} n \max_{k \neq l \in U} |\pi_{kl}-\pi_k \pi_l| < \infty$.
\end{itemize}
  Then (\ref{mean:square:cons}) holds and the Narain-Horvitz-Thompson estimator is consistent in mean square.
\end{prop}

\noindent The assumptions in Proposition \ref{prop0} are essentially the same as that in \citet{car:cha:gog:lab:2010}, except for the assumption (H3) which was replaced with
\begin{itemize}
  \item[] H3b. The variable $y$ is bounded, i.e. there exists some constant $C_1$ such that $|y_k| \leq C_1$.
\end{itemize}
\citet{car:gog:lar:2013} noticed however that (H3b) could be weakened to (H3). As noted by \citet{bre:ops:2000}, the assumption (H4) holds for stratified simple random sampling. This property also holds for rejective sampling \citep{haj:1964,boi:lop:rui:2012} and its Sampford-Durbin modification \citep{haj:1981}. However, this property is rather difficult to prove for other sampling designs with unequal probabilities. \\

\noindent When the variable $y$ has non-negative values, a first proposal is to replace (H4) with
    \begin{itemize}
      \item[] H4b: there exists some constant $a \geq 0$ such that for any vector $\pi$ of inclusion probabilities, we have for any $k \neq l \in U$:
        \begin{eqnarray}
          \pi_{kl} & \leq & \left(1+\frac{a}{n}\right) \times \pi_k \pi_l.
        \end{eqnarray}
    \end{itemize}
From (\ref{var:ht}), this leads to
    \begin{eqnarray} \label{var:ineg:2}
     V(\hat{t}_{y\pi}) & \leq & N^2 \left(\frac{1}{N~\min_{l \in U} \pi_l}+ \frac{a}{n} \right) \times \frac{1}{N} \sum_{k \in U} y_k^2.
    \end{eqnarray}

\begin{prop} \label{prop1}
  Assume that (H1)-(H3) and (H4b) hold, and that the variable $y$ has non-negative values. Then (\ref{mean:square:cons}) holds and the Narain-Horvitz-Thompson estimator is consistent in mean square.
\end{prop}

\noindent The assumption (H4) will hold in particular with $a=0$ when the sampling design satisfies the so-called Sen (1953)-Yates-Grundy (1953) conditions \nocite{sen:1953,yat:gru:1953}, namely $\pi_{kl} \leq \pi_k \pi_l$ for any $k \neq l \in U$. This property holds for stratified simple random sampling, and for several sampling algorithms with unequal probability such as Poisson sampling; the Midzuno method, the elimination method, Chao's method and the pivotal method \citep{dev:til:1998}; the Sampford design \citep{gab:1981,gab:1984}; the conditional Poisson sampling design \citep{che:dem:liu:1994}. \\
%This property will usually not be true for two-stage sampling, unless the first-stage sampling fraction is negligible. \\

\noindent In the case when the variable of interest may take both negative and non-negative values, we can consider the alternative condition that
    \begin{itemize}
      \item[] H4c: for any vector $\pi$ of inclusion probabilities, the variance of the Narain-Horvitz-Thompson estimator under the sampling design $p(\cdot)$ with parameter $\pi$ is no greater than the variance of the Hansen-Hurwitz estimator under multinomial sampling with parameter $\pi$.
    \end{itemize}
Under (H4c), it follows from (\ref{var:ht:mult}) that for any variable $y$
    \begin{eqnarray}
      V\left(\hat{t}_{y}\right) \leq \sum_{k \in U} \pi_k \left(\frac{y_k}{\pi_k}\right)^2 \leq \frac{N}{\min_{l \in U} \pi_l} \times \frac{1}{N} \sum_{k \in U} y_k^2. \label{var:ht:ineg2}
    \end{eqnarray}

\begin{prop} \label{prop2}
  Assume that (H1)-(H3) and (H4c) hold. Then (\ref{mean:square:cons}) holds and the Narain-Horvitz-Thompson estimator is consistent in mean square.
\end{prop}

\noindent The assumption (H4b) holds for simple random sampling, and for several sampling algorithms with unequal probability such as the Sampford design \citep{gab:1981,gab:1984}, the conditional Poisson sampling design \citep{qua:2008}, Chao's method \citep{sen:1989}, the elimination method \citep{dev:til:1998} and pivotal sampling \citep{cha:rui:2014}. Note that in case of pivotal sampling, numerous second-order inclusion probabilities are usually equal to zero \citep{dev:til:1998}, so that assumption (H4) does not hold while (H4b) and (H4c) are respected.

\section{Consistency for a martingale sampling algorithm} \label{sec:4}

\noindent A martingale sampling algorithm proceeds in steps $i=0,\ldots,T$ from $\pi(0)=\pi$ the vector of inclusion probabilities to $\pi(T)=I$ the final vector of sample membership indicators, such that the sequence $\{\pi(i)\}_{i=0,\ldots,T}$ is a discrete-time martingale with $\pi(i) \in [0,1]^N$ for any $i=0,\ldots,T$; see \citet{til:2011} and \citet{bre:cha:2011}. \\

\noindent Under a martingale sampling algorithm, we have $I-\pi=\sum_{i=0}^T \delta(i)$, where $\{\delta(i)\}_{i=0,\ldots,T}$ are the innovations of the martingale. Since these innovations are not correlated, we have
    \begin{eqnarray}
      V(I-\pi) = \sum_{i=0}^T V[\delta(i)] = E\left[ \sum_{i=0}^T \delta(i) \delta(i)^{\top} \right]. \label{var:mart:I}
    \end{eqnarray}
We can write $\hat{t}_y-t_y=\check{y}^{\top} (I-\pi)$ where $\check{y}=(\pi_1^{-1} y_1,\ldots,\pi_N^{-1} y_N)^{\top}$. From (\ref{var:mart:I}), we obtain
    \begin{eqnarray}
      V(\hat{t}_y-t_y) & = & E\left[\sum_{i=0}^T \sum_{k,l \in U} \frac{y_k}{\pi_k} \frac{y_l}{\pi_l} \delta_k(i) \delta_l(i) \right]. \label{var:mart:tyhat}
    \end{eqnarray}

\noindent We note
    \begin{eqnarray} \label{Ui}
      U_i & = & \{k \in U;~\delta_k(i) \neq 0\}
    \end{eqnarray}
the random subset of units in $U$ that are affected by step $i$. Also, we note $\mathcal{C}=\max_{i=0,\ldots,T} \textrm{Card}(U_i)$. From (\ref{var:mart:tyhat}), we obtain
    \begin{eqnarray}
      V(\hat{t}_y-t_y) & = & E\left[\sum_{i=0}^T \sum_{k,l \in U_i} \frac{y_k}{\pi_k} \frac{y_l}{\pi_l} \delta_k(i) \delta_l(i) \right] \nonumber \\
                       & \leq & E\left[\sum_{i=0}^T \sum_{k,l \in U_i} \left|\frac{y_k}{\pi_k}\right| \times \left|\frac{y_l}{\pi_l}\right| \right] \nonumber \\
                       & \leq & \left(\frac{\max_{k \in U} |y_k|}{\min_{k \in U} \pi_k} \right)^2 \times \mathcal{C}^2 \times E(T). \label{var:mart:tyhat2}
    \end{eqnarray}

\begin{prop} \label{prop3}
  Assume that assumptions (H1)-(H2) and (H3b) hold. Assume that $\mathcal{C}=O(1)$ and that $E(T)=O(N)$. Then (\ref{mean:square:cons}) holds and the Narain-Horvitz-Thompson estimator is consistent in mean square.
\end{prop}

\noindent Note that in Proposition \ref{prop3}, the stronger condition (H3b) on the variable $y$ is needed. Proposition \ref{prop3} is in particular useful when the sample $S$ is selected by means of the cube method \citep{dev:til:2004}. Suppose that a $q$-vector $x_k$ of auxiliary variables is known at the design stage for any unit $k \in U$. The $N \times q$ matrix $A = (x_k/\pi_k)_{k \in U}$ is called the matrix of constraints. The cube method enables to select samples such that the set of balancing equations
    \begin{eqnarray} \label{bal:samp}
      \sum_{k \in S} \frac{x_k}{\pi_k} & = & t_x
    \end{eqnarray}
is respected, at least approximately. A fast procedure for balanced sampling proposed by \citet{cha:til:2006,cha:til:2007} is described in Algorithm \ref{algo:1}. At any step $i$, $U_i \subset \{1,\ldots,N\}$ denotes the set of the $q+1$ first columns of $A$ such that $u_k(i)$ is not an integer. This is also the set of the $q+1$ first units in the population $U$ that are still neither selected nor rejected at step $i$. Also, $A_i$ denotes the sub-matrix of $A$ containing the columns in $U_i$. From the definition of $u(i)$ and $\delta(i)$ in Algorithm \ref{algo:1}, we have $\mathcal{C} \leq q+1$. Also, it is easily shown that $[(q+1)^{-1}N] \leq T \leq N$, with $[(q+1)^{-1}N]$ the largest integer smaller than $(q+1)^{-1}N$. Proposition \ref{prop4} below is thus an immediate consequence of Proposition \ref{prop3}.

\begin{algorithm}[htb!]
First initialize at $\pi(0)=\pi$. Next, at time $i=0,\cdots,T$, repeat the following steps:
\begin{enumerate}
  \item If there exists some vector $v(i) \neq 0$ such that $v(i) \in Ker(A_i)$, then:
    \begin{enumerate}
        \item Take any such vector $v(i)$ (random or not), and take $u(i)$ such that
            \begin{eqnarray*}
              u_k(i) = \left\{\begin{array}{ll}
                                               v_k(i) & \textrm{ if } k \in U_i,  \\
                                               0 & \textrm{ otherwise.}
                                             \end{array}
               \right.
            \end{eqnarray*}
        Compute $\lambda_1^*(i)$ and $\lambda_2^*(i)$, the largest values of $\lambda_1(i)$ and $\lambda_2(i)$ such that
            \begin{eqnarray*}
              0 \le \pi(i)+\lambda_1(i) u(i) \le 1 & \textrm{ and } & 0 \le \pi(i)-\lambda_2(i) u(i) \le 1.
            \end{eqnarray*}
        \item Take $\pi(i+1)=\pi(i)+\delta(i)$, where
            \begin{eqnarray*}
              \delta(i) = \left\{\begin{array}{ll}
                                               \lambda_1^*(i) u(i) & \textrm{ with probability } \lambda_2^*(i)/\lbrace \lambda_1^*(i)+\lambda_2^*(i) \rbrace,  \\
                                               -\lambda_2^*(i) u(i) & \textrm{ with probability } \lambda_1^*(i)/\lbrace \lambda_1^*(i)+\lambda_2^*(i) \rbrace.
                                             \end{array}
               \right.
            \end{eqnarray*}
    \end{enumerate}
  \item Otherwise, drop the last column from the matrix $A_i$ and go back to Step 1.
\end{enumerate}
\caption{A fast procedure for the cube method} \label{algo:1}
\end{algorithm}

\begin{prop} \label{prop4}
  Assume that assumptions (H1)-(H2) and (H3b) hold. Assume that the sample $S$ is selected by means of Algorithm \ref{algo:1}, and that $q=O(1)$. Then $V\left\{N^{-1}(\hat{t}_{y}-t_y)\right\}=O(n^{-1})$ and the Narain-Horvitz-Thompson estimator is consistent in mean square.
\end{prop}

\noindent Other implementations of the cube method are possible, for which Proposition \ref{prop3} may not be suitable to obtain the mean-square consistency. For the general balanced procedure described in Algorithm 8.3 in \citet{til:2011}, we have $U_i=\left\{k \in U;~\pi_k(i-1) \notin \{0,1\}\right\}$ which means that all the units that are still neither selected nor definitely rejected at step $i-1$ are possibly affected at step $i$. This leads to $\mathcal{C}=N$, so that the assumptions for Proposition \ref{prop3} are not fulfilled.

\bibliographystyle{apalike}
\bibliography{BiblioChauvet}

\begin{thebibliography}{}

\bibitem[Boistard et~al., 2012]{boi:lop:rui:2012}
Boistard, H., Lopuha{\"a}, H., and Ruiz-Gazen, A. (2012).
\newblock Approximation of rejective sampling inclusion probabilities and
  application to high order correlations.
\newblock {\em Electronic Journal of Statistics}, 6:1967--1983.

\bibitem[Breidt and Chauvet, 2011]{bre:cha:2011}
Breidt, F. and Chauvet, G. (2011).
\newblock Improved variance estimation for balanced samples drawn via the cube
  method.
\newblock {\em Journal of Statistical Planning and Inference}, 141(1):479--487.

\bibitem[Breidt and Opsomer, 2000]{bre:ops:2000}
Breidt, F. and Opsomer, J. (2000).
\newblock Local polynomial regression estimators in survey sampling.
\newblock {\em Annals of Statistics}, pages 1026--1053.

\bibitem[Cardot et~al., 2010]{car:cha:gog:lab:2010}
Cardot, H., Chaouch, M., Goga, C., and Labru{\`e}re, C. (2010).
\newblock Properties of design-based functional principal components analysis.
\newblock {\em Journal of Statistical Planning and Inference}, 140(1):75--91.

\bibitem[Cardot et~al., 2013]{car:gog:lar:2013}
Cardot, H., Goga, C., and Lardin, P. (2013).
\newblock Uniform convergence and asymptotic confidence bands for
  model-assisted estimators of the mean of sampled functional data.
\newblock {\em Electronic Journal of Statistics}, 7:562--596.

\bibitem[Chauvet and Ruiz-Gazen, 2014]{cha:rui:2014}
Chauvet, G. and Ruiz-Gazen, A. (2014).
\newblock A comparison of pivotal sampling and unequal probability sampling
  with replacement.
\newblock {\em Submitted}.

\bibitem[Chauvet and Till{\'e}, 2006]{cha:til:2006}
Chauvet, G. and Till{\'e}, Y. (2006).
\newblock A fast algorithm for balanced sampling.
\newblock {\em Computational Statistics}, 21(1):53--62.

\bibitem[Chauvet and Till{\'e}, 2007]{cha:til:2007}
Chauvet, G. and Till{\'e}, Y. (2007).
\newblock Application of fast sas macros for balancing samples to the selection
  of addresses.
\newblock {\em Case Studies in Business, Industry and Government Statistics},
  1:173--182.

\bibitem[Chen et~al., 1994]{che:dem:liu:1994}
Chen, X.-H., Dempster, A., and Liu, J. (1994).
\newblock Weighted finite population sampling to maximize entropy.
\newblock {\em Biometrika}, 81(3):457--469.

\bibitem[Deville and Till\'e, 1998]{dev:til:1998}
Deville, J.-C. and Till\'e, Y. (1998).
\newblock Unequal probability sampling without replacement through a splitting
  method.
\newblock {\em Biometrika}, 85(1):89--101.

\bibitem[Deville and Till\'e, 2004]{dev:til:2004}
Deville, J.-C. and Till\'e, Y. (2004).
\newblock Efficient balanced sampling: the cube method.
\newblock {\em Biometrika}, 91(4):893--912.

\bibitem[Gabler, 1981]{gab:1981}
Gabler, S. (1981).
\newblock A comparison of sampford's sampling procedure versus unequal
  probability sampling with replacement.
\newblock {\em Biometrika}, pages 725--727.

\bibitem[Gabler, 1984]{gab:1984}
Gabler, S. (1984).
\newblock On unequal probability sampling: sufficient conditions for the
  superiority of sampling without replacement.
\newblock {\em Biometrika}, 71(1):171--175.

\bibitem[H{\'a}jek, 1964]{haj:1964}
H{\'a}jek, J. (1964).
\newblock Asymptotic theory of rejective sampling with varying probabilities
  from a finite population.
\newblock {\em The Annals of Mathematical Statistics}, pages 1491--1523.

\bibitem[H{\'a}jek and Dupac, 1981]{haj:1981}
H{\'a}jek, J. and Dupac, V. (1981).
\newblock {\em Sampling from a finite population}.
\newblock Marcel Dekker New York.

\bibitem[Hansen and Hurwitz, 1953]{han:hur:1953}
Hansen, M.~H. and Hurwitz, W.~N. (1953).
\newblock Sample survey methods and theory. vol. i.

\bibitem[Horvitz and Thompson, 1952]{hor:tho:1952}
Horvitz, D. and Thompson, D. (1952).
\newblock A generalization of sampling without replacement from a finite
  universe.
\newblock {\em Journal of the American Statistical Association},
  47(260):663--685.

\bibitem[Isaki and Fuller, 1982]{isa:ful:1982}
Isaki, C. and Fuller, W. (1982).
\newblock Survey design under the regression superpopulation model.
\newblock {\em Journal of the American Statistical Association},
  77(377):89--96.

\bibitem[Narain, 1951]{nar:1951}
Narain, R. (1951).
\newblock On sampling without replacement with varying probabilities.
\newblock {\em Journal of the Indian Society of Agricultural Statistics},
  3:169--175.

\bibitem[Qualit{\'e}, 2008]{qua:2008}
Qualit{\'e}, L. (2008).
\newblock A comparison of conditional poisson sampling versus unequal
  probability sampling with replacement.
\newblock {\em Journal of Statistical Planning and Inference},
  138(5):1428--1432.

\bibitem[Robinson and S{\"a}rndal, 1983]{rob:sar:1983}
Robinson, P. and S{\"a}rndal, C.-E. (1983).
\newblock Asymptotic properties of the generalized regression estimator in
  probability sampling.
\newblock {\em Sankhy{\=a}: The Indian Journal of Statistics, Series B}, pages
  240--248.

\bibitem[Sen, 1953]{sen:1953}
Sen, A. (1953).
\newblock On the estimate of the variance in sampling with varying
  probabilities.
\newblock {\em Journal of the Indian Society of Agricultural Statistics},
  5(1194):127.

\bibitem[Sengupta, 1989]{sen:1989}
Sengupta, S. (1989).
\newblock On chao's unequal probability sampling plan.
\newblock {\em Biometrika}, 76(1):192--196.

\bibitem[Till{\'e}, 2011]{til:2011}
Till{\'e}, Y. (2011).
\newblock {\em Sampling algorithms}.
\newblock Springer.

\bibitem[Yates and Grundy, 1953]{yat:gru:1953}
Yates, F. and Grundy, P. (1953).
\newblock Selection without replacement from within strata with probability
  proportional to size.
\newblock {\em Journal of the Royal Statistical Society. Series B
  (Methodological)}, pages 253--261.

\end{thebibliography}

\end{document}